\documentclass[twocolumn]{autart}

\usepackage{amsmath,mathtools,amssymb}
\usepackage{color}
\usepackage{enumitem}
\usepackage{tikz}
\usetikzlibrary{arrows.meta}
\usetikzlibrary{automata, positioning}

\newtheorem{defin}{Definition}
\newtheorem{exam}{Example}
\newtheorem{assmpt}{Assumption}

\DeclareMathOperator*{\esssup}{ess\,sup}
\newcommand{\B}{\mathbb{B}}

\newcommand{\N}{\mathbb{N}}

\newcommand{\R}{\mathbb{R}}
\newcommand{\Rp}{\mathbb{R}_{\geq 0}}
\newcommand{\Rsp}{\mathbb{R}_{> 0}}

\newcommand{\cC}{\mathcal{C}}
\newcommand{\cD}{\mathcal{D}}

\newcommand{\cK}{\mathcal{K}}
\newcommand{\cL}{\mathcal{L}}

\newcommand{\cP}{\mathcal{P}}

\newcommand{\cS}{\mathcal{S}}
\newcommand{\cT}{\mathcal{T}}
\newcommand{\cU}{\mathcal{U}}
\newcommand{\cV}{\mathcal{V}}

\def\K{\mathcal{K}}
\def\Ki{\K_{\infty}}
\def\mer{\hfill$\circ$}

\def\iISS{\mathrm{iISS}}
\def\IPSS{\mathrm{IPSS}}
\def\id{\mathrm{id}}

\newcommand{\KL}{\mathcal{KL}}

\newcommand\norm[1]{\ensuremath{\lVert#1\rVert}}

\newcommand{\conc}[1]{\substack{\displaystyle\frown\\ ^{\scriptstyle #1}}}
\def\comp{{\scriptstyle{\,\circ\,}}} 

\newcommand{\hernan}[1]{#1} 

\renewcommand{\qed}{$\hfill\blacksquare$}
\allowdisplaybreaks

\begin{document}

\begin{frontmatter}
	\title{Input-Power-to-State Stability of Time-Varying Systems\thanksref{footnoteinfo}}
	
	\thanks[footnoteinfo]{Partially supported by Agencia I+D+i grant PICT 2021-I-A-0730, Argentina, and by the National Science Foundation of China under Grant 62203053. Corresponding author J. L. Mancilla-Aguilar.}
	
	\author[HHaddress]{Hernan Haimovich}\ead{haimovich@cifasis-conicet.gov.ar}, 
	\author[SLaddress]{Shenyu Liu}\ead{shenyuliu@bit.edu.cn}, 
	\author[ARaddress]{Antonio Russo}\ead{antonio.russo@unibg.it},
	\author[JLaddress]{Jose L. Mancilla-Aguilar}\ead{jmancil@fi.uba.ar}
	
	\address[HHaddress]{Centro Internacional Franco-Argentino de Cs. de la Informaci\'on y Sistemas (CIFASIS),\\ CONICET-UNR, Ocampo y Esmeralda, 2000 Rosario, Argentina.}
	\address[SLaddress]{School of Automation, Beijing Institute of Technology, 100081, Beijing, China.}
	\address[ARaddress]{Dipartimento di Ingegneria Gestionale, dell'Informazione e della Produzione, \\ Universit\`a degli Studi di Bergamo, 24044 Dalmine, Bergamo, Italy.}
	\address[JLaddress]{Departamento de Matem\'atica, Facultad de Ingenier\'{\i}a, Universidad de Buenos Aires, Av. Paseo Colon 850, CABA, Argentina.}
	
	
	\begin{abstract}
            When the state of a system may remain bounded even if both the input amplitude and energy are unbounded, then the state bounds given by the standard input-to-state stability (ISS) and integral-ISS (iISS) properties may provide no useful information. This paper considers an ISS-related concept suitable in such a case: input-power-to-state stability (IPSS). Necessary and sufficient conditions for IPSS are developed for time-varying systems under very mild assumptions on the dynamics. More precisely, it is shown that (a) the existence of a dissipation-form ISS-Lyapunov function implies IPSS, but not necessarily that of an implication-form one, (b) iISS with exponential class-$\KL$ function implies IPSS, and (c) ISS and stronger assumptions on the dynamics imply the existence of a dissipation-form ISS-Lyapunov function and hence IPSS. The latter result is based on a converse Lyapunov theorem for time-varying systems whose dynamics (i.e. state derivative) is not necessarily continuous with respect to time.
	\end{abstract}
	
\end{frontmatter}

\maketitle

\section{Introduction}

The state norm of an input-to-state stable (ISS) system is bounded by the sum of one term depending solely on the initial state norm and decaying to zero as time advances, and another term depending solely on the maximum input amplitude \cite{sontag_tac89,ISS:Mironchenko:2023}. Measuring instantaneous input amplitude amounts to the evaluation of some norm on the input-value space, a space which is usually finite-dimensional. Since all norms are equivalent in finite-dimensional spaces, then any choice of norm is suitable for the ISS property: the choice can only alter the specific bound on the state but not the fact that the system exhibits an ISS state-norm bound.

The input-related term in the state-norm bound provided by the related property of integral ISS (iISS) depends on the integral of some class $\K$ function of the input norm \cite{ISS:Sontag:1998,angson_tac00}. The latter function is called the iISS gain. The definition of the iISS property requires the existence of some iISS gain and the integral of the iISS gain evaluated on the input-value norm may be regarded as the energy provided by the input. Therefore, iISS can be interpreted as ISS with respect to input energy instead of amplitude (see  \cite{manroj_sicon23} for a general definition of ISS covering iISS as a special case). 


The ISS state-norm bound gives no useful information if the input amplitude does not have a finite bound, even if its integral remains bounded. In this case, the iISS state-norm bound may still provide some suitable information. However, if a system is such that the state may remain bounded even if both the input amplitude and energy become unbounded, then it is likely that neither ISS nor iISS will provide any useful information. One such case is when a system is subject to inputs with bounded average power, but a priori no amplitude or energy bounds, \hernan{as for example in spacecraft formation control \cite{gropan_ijrnc16}, where unbounded signals with bounded moving average may occur}. In these cases, it is useful to have some ISS-related property that bounds the state norm with respect to such average power or, equivalently, over the maximum energy over finite-time intervals of fixed length. Some existing works address this type of bound, such as \cite{gropan_ijrnc16,manhai_tac17,efifri_tac24}. Theorem~1 in \cite{gropan_ijrnc16} implies that if a time-invariant system is ISS, then some way of measuring input energy exists (given by a class $\Ki$ function) such that the state will eventually converge to any desired neighborhood of the origin for inputs with sufficiently small average power. By contrast, the way of measuring average input power in \cite[Theorem~4.1]{manhai_tac17} and \cite{efifri_tac24} are fixed, given by a moving average, and sufficient conditions are given for ISS to hold with respect to moving averages of the input. It is worthy of mention that the results of \cite{angnes_tac01}, although seemingly related, address a different type of concept because the bounds are not given on a norm of the state but on the average power of the state.



In this paper, we consider an ISS-related concept that bounds the state norm in terms of the initial state norm (and decays to zero) and a measure of average input power: the input-power-to-state stability (IPSS) property. As can be deduced directly from the definitions, the IPSS property implies both ISS and iISS and hence is not weaker than the combination of ISS and iISS. For time-invariant systems $\dot x = f(x,u)$ under a local Lipschitz continuity assumption on $f$, it is well-known that ISS implies iISS. For time-varying systems $\dot x = f(t,x,u)$, by contrast, systems with a locally Lipschitz continuous $f$ may be ISS and not iISS (see Proposition~2.9 in~\cite{haiman_auto19}) and hence not IPSS either. In addition, although for time-invariant systems under usual Lipschitz continuity assumptions the existence of an implication-form Lyapunov function is equivalent to the existence of a dissipation-form one, time-varying systems that are ISS may admit an implication-form ISS-Lyapunov function but no dissipation-form one. Two key questions then arise: (i) what stronger property is given by the existence of a dissipation-form ISS-Lyapunov function and (ii) whether IPSS is in fact \emph{equivalent} to the combination of ISS and iISS. In this context, the main contributions of the current paper are to show that for time-varying systems the existence of a dissipation-form ISS-Lyapunov function implies actually IPSS and to provide different sufficient conditions for IPSS to hold assuming that the system is either ISS or iISS. An important point is that the function $f$ defining the time-varying dynamics $\dot x = f(t,x,u)$ will not be assumed continuous with respect to $t$.
\section{Preliminaries}
\label{sec:preliminaries}

\subsection{Notation}
\label{sec:notation}
$\N$, $\R$, $\Rsp$ and $\Rp$ denote the natural numbers, reals, positive reals and nonnegative reals, respectively. We write $\alpha\in\cP$ if $\alpha:\R_{\ge 0} \to \R_{\ge 0}$ is continuous, $\alpha(0)=0$, and $\alpha(s) > 0$ for all $s>0$. We write $\alpha\in\K$ if $\alpha\in\cP$ and is strictly increasing, and $\alpha\in\Ki$ if $\alpha\in\K$ and is unbounded. We write $\beta\in\KL$ if $\beta:\R_{\ge 0}\times \R_{\ge 0}\to \R_{\ge 0}$, $\beta(\cdot,t)\in \K$ for any $t\ge 0$ and, for any fixed $r\ge 0$, $\beta(r,t)$ decreases to zero as $t\to \infty$. For any interval $I\subset\R$, the set $\cC^1(I)$ is the set of all continuously differentiable functions $\kappa : I \to \R$; if $\kappa \in \cC^1(I)$, then $\kappa'$ denotes its derivative. For any $p\in \N$, $|\cdot|$ denote the Euclidean norm in $\R^p$ and $\B^p_r=\{y\in \R^p:|y|\le r\}$ for every $r\ge 0$. Given $u:\Rp\to \R^m$, $v:\Rp\to \R^m$ and $\tau\ge 0$, $u\conc{\tau} v:\Rp \to \R^m$ is the concatenation of $u$ with $v$ at $\tau$, i.e. $u\conc{\tau} v(t)=u(t)$ if $t<\tau$ and $u\conc{\tau} v(t)=v(t)$ when $t\ge \tau$.  For a Lebesgue measurable function $u:\Rp\to \R^m$, $\|u\|_{\infty}=\esssup_{t\ge 0}|u(t)|$  denotes its essential supremum norm and, for a given $\rho \in \Ki$, $\|u\|_{\rho}=\int_0^{\infty}\rho(|u(s)|)ds$. Note that $\|\cdot\|_{\rho}$ is not necessarily a norm and that $\|u\|_{\rho}$ may equal $\infty$ for some $u$. For any interval $J\subset \Rp$, $u_J$ is the function that coincides with $u$ on $J$ and is $0$ elsewhere. 



 
\subsection{The time-varying system}
Consider the system
\begin{equation}\label{eq:f}
	\dot{x} = f(t,x,u)
\end{equation}
where $t\ge 0$, $x(t)\in \R^n$, $u(t)\in \R^m$ and $f \colon \Rp \times \R^n \times \R^m \to \R^n$. Along the paper, we consider the following standing assumption.
\begin{assmpt}
  \label{ass:standing}
  The function $f \colon \Rp \times \R^n \times \R^m \to \R^n$ satisfies
  \begin{enumerate}[label=S\arabic*)]
  \item $f(t,0,0)=0$ for all $t\ge 0$.\label{item:f00}
  \item $f(\cdot,\xi,\mu)$ is Lebesgue measurable for all $(\xi,\mu) \in \R^n \times \R^m$.\label{item:fmeast}
  \item $f(t,\cdot,\cdot)$ is continuous for all $t\geq 0$.\label{item:fcontxu}
  \item $f$ is bounded on bounded sets.\label{item:fbb}
  \end{enumerate}
\end{assmpt}
The name `input' is used to denote a function $u \colon \Rp \to \R^m$ that is Lebesgue measurable and locally essentially bounded, and $\cU$ denotes the set of all inputs. 
Assumption~\ref{ass:standing} ensures local existence of (possibly nonunique) solutions for every input and the boundedness implies continuation (BIC) property (\cite[p. 4]{karjia_book11}, \cite{karjia_ima11}): if a maximally defined solution is bounded, then it is defined for all future times. The set of all maximally defined solutions from initial conditions $(t_0,\xi) \in \R_{\ge 0} \times \R^n$ and corresponding to an input $u\in\cU$ is denoted by $\cS(t_0,\xi,u)$. The system~(\ref{eq:f}) is forward complete if for every $(t_0,\xi,u) \in \R_{\ge 0} \times \R^n \times \cU$, all solutions $x\in\cS(t_0,\xi,u)$ are defined for all $t\ge t_0$.
\subsection{Stability properties}

The \emph{uniform robust Lagrange stability} (URLS) \hernan{\cite[item P1 in Definition~2.3]{karjia_book11}}, \emph{uniform robust global asymptotic stability} (URGAS) \cite[Definition~2.3, Theorem~2.2]{karjia_book11}, \emph{input-to-state stability} (ISS) \cite{sontag_tac89} and \emph{integral input-to-state stability} (iISS) \cite{ISS:Sontag:1998} properties are next defined.  

\begin{defin} \label{def:ISS-iISS}
  A forward complete\footnote{The $\cV$-URLS and $\cV$-URGAS properties only require forward completeness for inputs in $\cV \subset \cU$.} system \eqref{eq:f} is 
  \begin{enumerate}
  \item $\cV$-URLS or URLS with respect to inputs in $\cV \subset \cU$ if for every $\delta>0$ there exists $\varepsilon = \varepsilon(\delta) > 0$ such that for all $t_0\ge 0$, $\xi \in \B_\delta^n$, $u\in \cV$ and $x\in\cS(t_0, \xi, u)$,
  \begin{equation}
      \label{eq:URLS}
      \sup_{t\ge t_0} |x(t)| \le \varepsilon;
  \end{equation}
  \item $\cV$-URGAS or URGAS with respect to inputs in $\cV \subset \cU$ if there exists $\beta\in \KL$ so that for all $t_0\ge 0$, $\xi \in \R^n$, $u\in \cV$ and $x\in\cS(t_0, \xi, u)$,
    \begin{equation}
      \label{eq:RGAS}
      |x(t)|\leq \beta(|\xi|,t-t_0) \quad \forall t\geq t_0;
    \end{equation}
  \item ISS if there exist $\beta\in \KL$ and $\eta \in \Ki$ so that for all $t_0\ge 0$, $\xi \in \R^n$, $u\in \cU$ and $x\in\cS(t_0, \xi, u)$,
    \begin{equation}\label{ISS}
      |x(t)|\leq \beta(|\xi|,t-t_0) + \eta(\norm{u}_{\infty}) \quad \forall t\geq t_0;
    \end{equation}
  \item iISS if there exist $\beta\in \KL$ and $\rho,\gamma \in \Ki$ so that for all $t_0\ge 0$, $\xi \in \R^n$, $u\in \cU$ and $x\in\cS(t_0, \xi, u)$,
    \begin{equation}\label{iISS}
      |x(t)|\leq \beta(|\xi|,t-t_0) + \gamma(\norm{u}_{\rho}) \quad \forall t\geq t_0.
    \end{equation}
  \end{enumerate}
\end{defin}
The state of a system satisfying the $\cV$-URLS property remains bounded uniformly with respect to initial time, all inputs in $\cV$, and all initial states with bounded norm. The $\cV$-URGAS property bounds the state norm with a convergent function depending on the initial state, irrespective of which input in $\cV$ is applied. If $\cV$ contains only the zero input, then $\cV$-URGAS becomes the usual 0-GUAS and $\cV$-URLS becomes Lagrange stability uniformly with respect to initial time. The ISS property bounds the state norm in terms of the initial state and the maximum amplitude of the input signal. Interpreting $\|u\|_{\rho}$ as the energy of $u$, iISS bounds the state norm in terms of the initial condition and the input energy. All of these stability properties are uniform with respect to initial time, meaning that the relationship $\varepsilon=\varepsilon(\delta)$ and the comparison functions $\beta,\eta,\gamma,\rho$ are the same for different values of $t_0$.

Some systems may exhibit bounded states even under inputs with unbounded amplitude and energy. One such class of inputs is that with bounded average power.
Given $\rho \in \cK_\infty$ and $T>0$, for $u\in\cU$ let
\begin{equation}\label{def:power_norm}
	\norm{u}_{\rho,T} := \frac{1}{T} \sup\limits_{t\geq 0} \int_{\max\{t-T,0\}}^{t} \rho (|u(s)|) \, ds.
\end{equation}
The value $\| u \|_{\rho,T}$ can be interpreted as the input's maximum average power over time intervals of length $T$. 

The following stability property bounds the state norm in relation to the input's maximum average power.
\begin{defin}
  System \eqref{eq:f} is input-power-to-state stable (IPSS) if it is forward complete and there exist $\beta \in \cK\cL$, $\gamma\,,\rho \in \cK_\infty$ and $T>0$ so that for all $t_0 \geq 0$, $\xi\in\R^n$, $u\in\cU$ and $x\in\cS(t_0, \xi, u)$, 
  \begin{equation}\label{IPSS}
    |x(t)|\leq \beta(|\xi|,t-t_0) + \gamma(\norm{u}_{\rho,T}) \quad \forall t\geq t_0.
  \end{equation}
\end{defin}

Note that $\|\cdot\|_{\rho,T}$ is not necessarily a norm and that 
for any $T,T^*>0$, there exist constants $k,k^*>0$ so that $\|u\|_{\rho,T}\le k\|u\|_{\rho,T^*}$ and $\|u\|_{\rho,T^*}\le k^*\|u\|_{\rho,T}$ \cite[Proposition 2.2]{manhai_scl17}. Hence, the definition of IPSS is actually independent of the number $T$.
\begin{rem} \label{rem:IPSS}
  By causality and the semigroup (Markov) property of solutions, the input $u$ can be replaced by 
  any function coinciding with $u$ within $[t_0,t]$ 
  in any of the bounds (\ref{ISS}), (\ref{iISS}) or (\ref{IPSS}), yielding equivalent definitions.\mer
\end{rem}

IPSS is easily seen to be stronger than ISS and iISS, as established in the following lemma. 
\begin{lem}
  \label{lem:ipss-stronger}
  If system (\ref{eq:f}) is IPSS then it is ISS and iISS.
\end{lem}
\begin{pf}
  Let $u \in \cU$. Then, $\|u\|_{\rho,T}\le \rho(\|u\|_{\infty})$ and $\|u\|_{\rho,T}\le \|u\|_{\rho}/T$. As a consequence, (\ref{IPSS}) implies (\ref{ISS}), and also (\ref{iISS}) with $\gamma_{\iISS}(\cdot) = \gamma_{\IPSS}(\,\cdot\,/T)$.\qed
\end{pf}
\begin{rem}
    A subscript $_\iISS$ or $_\IPSS$ is added to the functions $\gamma$ in \eqref{iISS} or \eqref{IPSS} to distinguish one from the other. \mer
\end{rem}
\hernan{%
\begin{exam}
    Let $\tau\ge 1$ and consider a system~\eqref{eq:f} that is IPSS with gain $\rho(s) = \sqrt{s}$, and hence also ISS and iISS. Consider an input $u : \R_{\ge 0} \to \R^m$ satisfying
    \begin{align*}
    |u(t)| \in 
    \begin{cases}
        [k^2,2k^2] &\text{if }t\in [k\tau,k\tau+\frac{1}{k}), k\in\N, \\
        \{0\} &\text{otherwise.}        
    \end{cases} 
    \end{align*}
    This input satisfies (cf. \cite[Example~1]{gropan_ijrnc16})
    \begin{align*}
        \norm{u}_{\infty} &= \infty, &
        \norm{u}_{\rho} &= \infty, &
        \norm{u}_{\rho,T} &< \infty, &
        \norm{u}_{\id,T} &= \infty,        
    \end{align*}
    for all $T>0$, with $\rho(s) = \sqrt{s}$ for all $s\ge 0$. Therefore, the bounds \eqref{ISS} and \eqref{iISS} are unable to provide useful information as $t\to\infty$, whereas \eqref{IPSS} does provide a finite bound. Moreover, since the linear moving average $\norm{u}_{\id,T}$ is also infinite, then the results in \cite{efifri_tac24} are also uninformative.
\end{exam}
}

\subsection{Lyapunov functions}\label{subsec:V}

Lyapunov functions are an essential tool for establishing stability properties without computing solutions. Weaker versions of the standard ISS- and iISS-Lyapunov functions \cite{sontag_tac89,ISS:Sontag:1998} are next defined, where only Lipschitz continuity instead of smoothness is required. This relaxation is suitable in the current time-varying context where $f$ in~(\ref{eq:f}) is not necessarily continuous with respect to $t$.
\begin{defin}
  \label{def:Lyap}
  A locally Lipschitz function $V\colon \R_{\geq 0} \times \R^n \to \R$ for which there exist $\alpha_1, \alpha_2 \in \cK_\infty$ such that 
  \begin{equation}
    \label{eq:sandwich}
    \alpha_1(|\xi|) \leq V(t,\xi) \leq \alpha_2(|\xi|)\quad \forall (t,\xi)\in\R_{\ge 0} \times \R^n,
  \end{equation}
  is said to be
  \begin{itemize}
  \item an implication-form ISS-Lyapunov function for system~(\ref{eq:f}) if there exist $\alpha_3\in\cP, \chi_3\in\Ki$ and a zero-measure set $\cT \subset \R_{\ge 0}$ such that 
    \begin{align}
      \label{eq:impform}
      D^+_f V(t,\xi,\mu) &\le -\alpha_3(|\xi|)
    \end{align}
    holds for every $(t,\xi,\mu) \in \R_{\ge 0}\setminus\cT \times \R^n \times \R^m$ for which $|\xi| \ge \chi_3(|\mu|)$;
  \item a dissipation-form ISS-Lyapunov function for system~(\ref{eq:f}) if there exist $\alpha_4,\chi_4\in\Ki$ and a zero-measure set $\cT \subset \R_{\ge 0}$ such that 
    \begin{align}
      \label{eq:dissipform}
      D^+_f V(t,\xi,\mu) &\le -\alpha_4(|\xi|) + \chi_4(|\mu|)
    \end{align}
    holds for all $(t,\xi,\mu) \in \R_{\ge 0}\setminus\cT \times \R^n \times \R^m$;
  \item an iISS-Lyapunov function for system~(\ref{eq:f}) if there exist $\alpha_5\in \cP$, $\chi_5\in\Ki$ and a zero-measure set $\cT \subset \R_{\ge 0}$ such that 
    \begin{align}
      \label{eq:iiss-Lyap}
      D^+_f V(t,\xi,\mu) &\le -\alpha_5(|\xi|) + \chi_5(|\mu|)
    \end{align}
    holds for all $(t,\xi,\mu) \in \R_{\ge 0}\setminus\cT \times \R^n \times \R^m$;
  \end{itemize}
  with $D^+_f V$ defined as
  \begin{multline}
    \label{eq:Dpf}
    D^+_f V(t,\xi,\mu) := \\
    \limsup_{h\to 0^+} \frac{V(t+h,\xi + hf(t,\xi,\mu))-V(t,\xi)}{h}.
  \end{multline}
\end{defin}
\hernan{If $V$ is continuously differentiable, then the derivative~\eqref{eq:Dpf} becomes the usual $\frac{\partial V}{\partial t} + \frac{\partial V}{\partial x}f$.}
If $V$ is a dissipation-form ISS-Lyapunov function, it is easily seen that it is also an implication-form one. This follows defining $\chi_3\in\Ki$ via $\chi_3(s) = \alpha_4^{-1}(2\chi_4(s))$, leading to $\alpha_3 \in \Ki \subset \cP$ satisfying $\alpha_3(s) = \alpha_4(s)/2$ in~\eqref{eq:impform}.
The existence of a Lyapunov-type function as per Definition~\ref{def:Lyap} is sufficient to ensure the corresponding stability property.
\begin{prop}
    \label{prop:Lyapunov}
  Let $f:\R_{\ge0}\times\R^n\times\R^m$ satisfy Assumption~\ref{ass:standing} and let $V : \R_{\ge0}\times\R^n \to \R_{\ge0}$. If $V$ is a (dissipation- or implication-form) ISS-Lyapunov function, then~(\ref{eq:f}) is ISS. If $V$ is an iISS-Lyapunov function, then~(\ref{eq:f}) is iISS.
\end{prop}
\begin{pf}
    The parts of the proof that may differ with respect to the time-invariant, smooth Lyapunov function case are explained.
    Let $t_0 \in \Rp$, $\xi_0\in\R^n$, $u\in\cU$ and consider a solution $x \in \cS(t_0,\xi_0,u)$, defined on $[t_0,t_x)$. Define $v : [t_0,t_x) \to \Rp$ via $v(t) = V(t,x(t))$. Since $V$ is locally Lipschitz and $x$ is locally absolutely continuous on $(t_0,t_x)$, then $v$ is also locally absolutely continuous and hence differentiable almost everywhere within $(t_0,t_x)$. Therefore
    \begin{align*}
        \dot v(t) &= \lim_{h\to 0}\frac{v(t+h)-v(t)}{h} = 
        \limsup_{h\to 0^+} \frac{v(t+h)-v(t)}{h}
    \end{align*}
    for almost all $t\in (t_0,t_x)$. 
    \hernan{Since $V$ is locally Lipschitz and $x$ satisfies~\eqref{eq:f} almost everywhere, then 
    \begin{align*}
        \limsup_{h\to 0^+} \frac{v(t+h)-v(t)}{h} = D^+_fV(t,x(t),u(t))
    \end{align*}
    for almost all $t\in (t_0,t_x)$ follows after adding and subtracting $V(t+h,x(t)+h \dot x(t))$ in the numerator and using the Lipschitz bound on $V$, the definition of derivative, and~\eqref{eq:f}.} In consequence, there exists a zero-measure set $\cT_1 \subset (t_0,t_x)$ such that for all $t\in (t_0,t_x)\setminus \cT_1$
    \begin{align*}
        \dot{v}(t) = D^+_fV(t,x(t),u(t)).
    \end{align*}
    For every $t\ge t_0$, define
    \begin{align*}
        U_t := \esssup_{t_0 \le s < t} |u(s)| = \norm{u_{[t_0,t)}}_{\infty}.
    \end{align*}
    Then,
    \begin{enumerate}[wide, labelindent=0pt]
        \item[(i)] if $V$ is a dissipation-form ISS-Lyapunov function, then it also is an implication-form one,
        \item[(ii)] if $V$ is an implication-form ISS-Lyapunov function, then
        \begin{align*}
            \dot v(t) = D_f^+ V(t,x(t),u(t)) \le -\alpha_3(|x(t)|)
        \end{align*}
        holds for all $t\in (t_0,t_x) \setminus (\cT_1 \cup \cT)$ and provided that $|x(t)| \ge \chi_3(U_{t_x})$.
        Analogously to the time-invariant case, using the bounds~\eqref{eq:sandwich}, it can be shown that there exists a locally Lipschitz $\sigma_3\in\cP$ such that $\alpha_3(|\xi|) \ge \sigma_3(V(t,\xi))$ holds for all $(t,\xi) \in \Rp \times \R^n$. Then,
        \begin{align*}
            \dot v(t) = D_f^+ V(t,x(t),u(t)) \le -\sigma_3(v(t))
        \end{align*}
        for almost all $t\in (t_0,t_x)$ and provided that $v(t) \ge \alpha_2\comp \chi_3(U_{t_x})$. 
        From here on, the arguments are analogous to the smooth or time-invariant cases, showing that the solution must remain bounded, hence $t_x = \infty$, and that the ISS bound~\eqref{ISS} can be obtained (see e.g. Theorems~4.9, 4.18 and 4.19 of \cite{khalil_book02} with differentiability replaced by almost everywhere differentiability and the corresponding comparison Lemma~3.4 of \cite{khalil_book02} by \cite[Corollary~IV.3]{angson_tac00}).
        \item[(iii)] If $V$ is an iISS-Lyapunov function, then 
        \begin{align*}
            \dot v(t) = D_f^+ V(t,x(t),u(t)) &\le -\alpha_5(|x(t)|) + \chi_5(|u(t)|) \\
            &\le -\sigma_5(v(t)) + \chi_5(|u(t)|)
        \end{align*}
        holds for all $t\in (t_0,t_x) \setminus (\cT_1  \cup \cT)$ for some locally Lipschitz $\sigma_5\in\cP$. From here on, the arguments are the same as for the smooth or time-invariant case, showing that the solution must remain bounded if $\|u_{[t_0,t_x)}\|_{\chi_5}$ is finite, and hence $t_x = \infty$, and that the iISS bound with $\rho = \chi_5$ can be obtained \cite[Proof of 2$\Rightarrow$1 of Theorem 1]{angson_tac00}.
        \qed
    \end{enumerate}
\end{pf}
Since $\Ki\subset\cP$, any dissipation-form ISS-Lyapunov function is also an iISS-Lyapunov function. For time-invariant systems under standard continuity and boundedness assumptions on $f$, it is well-known that the existence of an implication-form ISS-Lyapunov function is equivalent to that of a dissipation-form one \cite{sonwan_scl95}, the existence of an iISS-Lyapunov function is equivalent to iISS, and ISS implies iISS \cite{angson_tac00}. For time-varying systems, by contrast, ISS systems satisfying Assumption~\ref{ass:standing} and having an implication-form ISS-Lyapunov function but no dissipation-form one exist \cite{edwlin_cdc00}. In addition, a time-varying system that satisfies Assumption~\ref{ass:standing} can be ISS and not iISS \cite[Proposition~2.9]{haiman_auto19}. 

\section{Sufficient Lyapunov condition for IPSS}

Given that the existence of a dissipation-form ISS-Lyapunov function implies that of an implication-form one but not conversely, and that an implication-form one is sufficient to ensure ISS, a natural question is what stronger property does the existence of a dissipation-form ISS-Lyapunov function ensure in the time-varying setting. An answer to this question is one of the main results of this paper, namely that a dissipation-form ISS-Lyapunov function implies IPSS and not just ISS (recall Lemma~\ref{lem:ipss-stronger}).

\begin{thm}
    \label{thm:Lyap-ipss}
  Let Assumption~\ref{ass:standing} hold and let $V : \R_{\ge0}\times\R^n \to \R_{\ge0}$ be a dissipation-form ISS-Lyapunov function. Then, system \eqref{eq:f} is IPSS.
\end{thm}
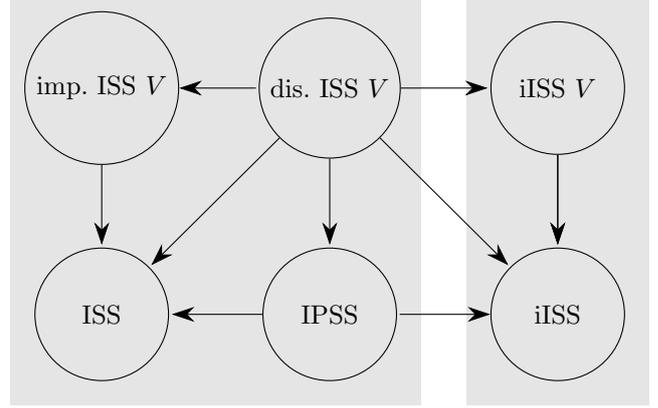
\begin{figure}
    \centering
    \begin{tikzpicture}
        \fill [gray, opacity=0.2] (-1.2,-1.2) rectangle (4.2,4.2);
        \fill [gray, opacity=0.2] (4.8,-1.2) rectangle (7.2,4.2);
	\node[state, minimum size=5em] at (0, 3)  (ifISSV) {imp. ISS $V$};
	\node[state, minimum size=5em] at (3, 3)  (dfISSV) {dis. ISS $V$};
	\node[state, minimum size=5em] at (6, 3)  (iISSV) {iISS $V$};
	\node[state, minimum size=5em] at (0, 0)  (ISS) {ISS};
	\node[state, minimum size=5em] at (3, 0)  (IPSS) {IPSS};  
 	\node[state, minimum size=5em] at (6, 0)  (iISS) {iISS};  
	\draw[every loop]
	(ifISSV) edge[-{Stealth[length=3mm,width=2mm]}] (ISS)
	(dfISSV) edge[-{Stealth[length=3mm,width=2mm]}] (IPSS)
	(iISSV) edge[-{Stealth[length=3mm,width=2mm]}] (iISS)
	(ifISSV) edge[{Stealth[length=3mm,width=2mm]}-] (dfISSV)
	(dfISSV) edge[-{Stealth[length=3mm,width=2mm]}] (iISSV)
	(iISS) edge[{Stealth[length=3mm,width=2mm]}-] (IPSS)
	(IPSS) edge[-{Stealth[length=3mm,width=2mm]}] (ISS)
        (iISSV) edge[-{Stealth[length=3mm,width=2mm]}] (iISS)
        (dfISSV) edge[-{Stealth[length=3mm,width=2mm]}] (ISS)
        (dfISSV) edge[-{Stealth[length=3mm,width=2mm]}] (iISS);
\end{tikzpicture}
    \caption{Relations between the existence of implication-form ISS-Lyapunov function, existence of dissipation-form ISS-Lyapunov function, existence of iISS-Lyapunov function, ISS, IPSS, and iISS. Under Assumption~\ref{ass:standing}, only the implications marked by arrows are known. Additionally when $f$ is independent of $t$ and satisfies continuity and boundedness assumptions, then the properties in each connected shaded area are equivalent.}
    \label{fig:1}
\end{figure}
%
%
%
\begin{pf}
    By Proposition~\ref{prop:Lyapunov}, system~\eqref{eq:f} is ISS and hence forward complete. 
  Let $\alpha_1,\alpha_2,\alpha_4,\chi_4$ characterize the dissipation-form ISS-Lyapunov function, as per Definition~\ref{def:Lyap}.
  For $\kappa \in \cK_\infty \cap \cC^1(\R_{\geq 0})$ to be selected later, define $W := \kappa \comp V$. Then, \eqref{eq:sandwich} implies that
  \begin{equation}
    \label{eq:new_sandwich}
    \alpha_2^{-1}\comp\kappa^{-1}(W(t,\xi))\leq |\xi|\leq \alpha_1^{-1}\comp\kappa^{-1}(W(t,\xi)),
  \end{equation}
  for all $(t,\xi) \in \R_{\ge 0}\times \R^n$. Let $t_0\ge 0$, $\xi_0 \in \R^n$ and $u\in \cU$, and consider any solution $x\in \cS(t_0,\xi_0,u)$. Define $w(t):=W(t,x(t))$ and $v(t)=V(t,x(t))$.
 Since $\kappa$ is continuously differentiable, $V$ is locally Lipschitz, and $x$ is locally absolutely continuous and satisfies~\eqref{eq:f} for almost all $t\ge t_0$, by using the same arguments as in the proof of Proposition \ref{prop:Lyapunov} and that $V$ is a dissipation-form ISS, we have that
  \begin{align*}
    \dot w(t) &= \kappa'(v(t))\cdot \dot{v}(t)=\kappa'(v(t)) \cdot D^+_f V(t,x(t),u(t))\\
    &\le \kappa'(v(t)) [-\alpha_4(|x(t)|) + \chi_4(|u(t)|)]
  \end{align*}
  holds for almost all $t\ge t_0$. Similarly to the construction in \cite[Lemmas~11, 12]{prawan_mcss96}, let $\kappa$ be defined via
  \begin{align*}
      \kappa(q) &:= \exp\left(2\int_1^q \frac{d\tau}{a(\tau)} \right),\\
      a(\tau) &:= \frac{2}{\pi}\int_0^\tau \frac{\min\{s,\sigma(s)\}}{1+s^2} ds,\\
      \hernan{\sigma} &:= \alpha_4\comp\alpha_2^{-1} \in \Ki.
  \end{align*}
  Then, $\kappa \in \cK_\infty \cap \cC^1(\R_{\geq 0})$, $\kappa'$ is nondecreasing and nonnegative, and
  \begin{align*}
      \kappa'(s) \sigma(s) &\ge 2\kappa(s) \quad \forall s\ge 0.
  \end{align*}
  It follows that 
  \begin{align*}
      \dot w(t) &\le \kappa'(v(t)) [-\alpha_4(|x(t)|) + \chi_4(|u(t)|)]\\
      &\le \kappa'(v(t)) [-\sigma(v(t)) + \chi_4(|u(t)|)]\\
      &\le -\kappa(v(t)) +
     \kappa'(v(t))[-\sigma(v(t))/2 + \chi_4(|u(t)|)]
  \end{align*}
  for almost all $t\ge t_0$.
  Following the steps in the proof of (b) $\Rightarrow$ (c) of Theorem 2 in \cite{heslib_auto08}, then 
  \begin{equation}\label{exp_W}
    \dot{w}(t) \leq -w(t)+\rho(|u(t)|)  
  \end{equation}
  for almost all $t\ge t_0$, with $\rho\in\Ki$ defined via $\rho(s)=\kappa'\comp \sigma^{-1} (2\chi_4(s)) \chi_4(s)$.
  From the comparison principle, for any $t\geq t_0\geq 0$, then
  \begin{equation}\label{W_bound}
    w(t)\leq e^{-(t-t_0)}w(t_0)+\int_{t_0}^te^{-(t-\tau)}\rho(|u(\tau)|)d\tau.
  \end{equation}
  Pick $T>0$ and define $N=N(t_0,t):=\lfloor\frac{t-t_0}{T}\rfloor+1$. It holds that $t_0+(N-1)T\leq t<t_0+NT$, and 
  \begin{align*}
    \int_{t_0}^te^{-(t-\tau)}&\rho(|u(\tau)|)d\tau \leq \int_{t_0}^{t_0+NT}e^{-(t-\tau)}\rho(|u(\tau)|)d\tau \\
    & \leq \int_{t_0}^{t_0+NT}e^{-(t_0+(N-1)T-\tau)}\rho(|u(\tau)|)d\tau \\
    &\leq \sum_{i=0}^{N-1}\int_{t_0+iT}^{t_0+(i+1)T}e^{-(N-i-2)T}\rho(|u(\tau)|)d\tau\\
    &=\sum_{i=0}^{N-1}e^{-(N-i-2)T}\int_{t_0+iT}^{t_0+(i+1)T}\rho(|u(\tau)|)d\tau.
  \end{align*}
  From \eqref{def:power_norm}, we have $\int_{t_0+iT}^{t_0+(i+1)T}\rho(|u(\tau) |)d\tau\leq T \Vert u\Vert_{\rho,T}$ for all $i=0,1,\ldots,N-1$. Thus,
  \begin{align}\label{bound_on_u}
    \int_{t_0}^te^{-(t-\tau)}\rho(|u(\tau)|)d\tau &\leq \sum_{i=0}^{N-1}e^{-(N-i-2)T} T \Vert u\Vert_{\rho,T}\nonumber \\
    &=\frac{e^{2T}-e^{-(N-2)T}}{e^T-1} T \Vert u\Vert_{\rho,T} \nonumber \\
    &\leq \frac{e^T}{1-e^{-T}} T \Vert u\Vert_{\rho,T},
  \end{align}
  which is independent of $N$. Substitute \eqref{bound_on_u} into \eqref{W_bound} and use the bounds \eqref{eq:new_sandwich} so that
  \begin{align*}
    |x(t)|&\leq \alpha_1^{-1}\comp\kappa^{-1}\big(w(t)\big)\\
    &\leq \alpha_1^{-1}\comp\kappa^{-1}\left(e^{-(t-t_0)}w(t_0)+\frac{e^T T}{1-e^{-T}}\Vert u\Vert_{\rho,T}\right)\\
    &\leq \alpha_1^{-1}\comp\kappa^{-1}\left(2e^{-(t-t_0)}w(t_0)\right)\\
    &+\alpha_1^{-1}\comp\kappa^{-1}\left(\frac{2e^T T}{1-e^{-T}}\Vert u\Vert_{\rho,T}\right)\\
    &\leq \alpha_1^{-1}\comp\kappa^{-1}\left(2e^{-(t-t_0)}\kappa\comp\alpha_2(|x(t_0)|)\right)\\
    &+\alpha_1^{-1}\comp\kappa^{-1}\left(\frac{2e^T T}{1-e^{-T}}\Vert u\Vert_{\rho,T}\right).
  \end{align*}
  This leads to the estimate \eqref{IPSS} with $\beta\in\KL$ and $\gamma\in \Ki$ defined via $\beta(s,t):=\alpha_1^{-1}\comp\kappa^{-1}\left(2e^{-t}\cdot\kappa\comp\alpha_2(s)\right)$ and $\gamma(s):=\alpha_1^{-1}\comp\kappa^{-1}\left(\frac{2e^T T s}{1-e^{-T}}\right)$. \qed 
\end{pf}
			



The implications given by Lemma~\ref{lem:ipss-stronger}, Proposition~\ref{prop:Lyapunov}, and Theorem~\ref{thm:Lyap-ipss} are summarized in Fig.~\ref{fig:1}.
\hernan{An interesting question is whether the converse of Theorem~\ref{thm:Lyap-ipss} also holds. An answer to this question under stronger assumptions is given by Theorem~\ref{thm:ISS_to_IPSS} in Section~\ref{sec:IPSS-from-ISS}.}

\section{When ISS or iISS imply IPSS}

In general, neither ISS nor iISS implies IPSS. This follows from the existence of systems that are iISS and not ISS \cite{angson_tac00}, systems that are ISS and not iISS \cite[Proposition~2.9]{haiman_auto19}, and the fact that IPSS implies both ISS and iISS (recall Lemma~\ref{lem:ipss-stronger}). 


\subsection{Sufficient condition for IPSS from iISS}

A simple sufficient condition for an iISS system to be IPSS is that its $\KL$-function $\beta$ be exponential.
\begin{prop}
  \label{prop:betaexp}
  Let system~(\ref{eq:f}) be iISS, satisfying~(\ref{iISS}) with $\beta(r,t) = K r e^{-\lambda t}$ for some $K,\lambda > 0$. Then, $K\ge 1$ and (\ref{eq:f}) is IPSS and satisfies~(\ref{IPSS}) with $T> \log(K)/\lambda$, $\beta(r,t) = K r e^{-\tilde\lambda t}$, $\tilde\lambda$ as in~(\ref{eq:tildelambda}) and $\gamma_{\IPSS} = T\frac{1+K(1-e^{-\lambda T})}{1-Ke^{-\lambda T}} \gamma_{\iISS}$. 
\end{prop}
The proof of Proposition~\ref{prop:betaexp} follows from application of the following lemma.
\begin{lem}
  \label{lem:exponential}
  Let $g,h,\eta:\R_{\ge0}\to\R_{\ge0}$ with $\eta$ nondecreasing, and let $K,\lambda >0$. If
  \begin{align}
    \label{eq:exp-1}
    g(t) &\le K g(t_0) e^{-\lambda (t-t_0)} + \eta\left(\int_{t_0}^t h(\tau)d\tau\right) 
  \end{align}
  for all $t\ge t_0 \ge 0$, then the following inequality holds for every $T> \frac{1}{\lambda}\log(\max\{1,K\})$ and for all $t\ge t_0 \ge 0$: 
  \begin{align}
    \label{eq:exp-2}
    g(t) &\le K g(t_0) e^{-\tilde\lambda (t-t_0)} + \frac{1 + K (1-e^{-\lambda T})}{1-Ke^{-\lambda T}} \eta(\phi) \\
    \label{eq:tildelambda}
    \tilde\lambda &:= \lambda - \frac{\log(\max\{1,K\})}{T},\\
    \phi &:= \sup_{s\in [t_0,t]} \int_{\max\{t_0,s-T\}}^s h(\tau)d\tau.\notag
  \end{align}
\end{lem}

\begin{pf}(Proposition~\ref{prop:betaexp})
  Apply Lemma~\ref{lem:exponential} to the inequality~(\ref{iISS}) setting $g(t) = |x(t)|$, $h(t) = \rho(|u(t)|)$ and $\eta=\gamma$.\qed
\end{pf}

\begin{pf}(Lemma~\ref{lem:exponential})
  Let $T> \frac{1}{\lambda}\log(\max\{1,K\})$, $t \ge t_0 \ge 0$, and define $t_\ell = \ell T + t_0$ for all $\ell\in\N$. Let $k\in\N$ be such that $t \in [t_k,t_{k+1})$. From~(\ref{eq:exp-1}), then
    \begin{align*}
      g(t) &\le K g(t_k) e^{-\lambda(t-t_k)} + \eta\left(\int_{t_k}^t h(\tau)d\tau\right) \\
      &\le K g(t_k) e^{-\lambda(t-t_k)} + \eta(\phi),\\
      g(t_\ell) &\le K g(t_{\ell-1}) e^{-\lambda T} + \eta\left(\int_{t_{\ell-1}}^{t_\ell} h(\tau)d\tau\right)\\
      &\le K g(t_{\ell-1}) e^{-\lambda T} + \eta(\phi) 
    \end{align*}
    where the second and last inequalities follow from $\eta$ being nondecreasing.
    From the inequality satisfied by $T$, then $Ke^{-\lambda T} =: C < 1$. Define $\tilde\lambda = \lambda - \log(\max\{1,K\})/T > 0$.
    Iterating the last inequality, then
    \begin{align*}
      &g(t_\ell)
      \le  C^\ell g(t_0) + \eta(\phi) \sum_{p=0}^{\ell-1} C^p 
      \le C^\ell g(t_0) + \eta(\phi)\frac{1- C^\ell}{1-C}\\
      &\le e^{-\tilde\lambda (t_\ell - t_0) } g(t_0) + \eta(\phi) \frac{1- C^\ell}{1-C}
    \end{align*}
  Setting $\ell=k$ and substituting into the inequality for $g(t)$, it follows that
  \begin{align*}
    g(t) &\le K \left( e^{-\tilde\lambda (t_k - t_0) } g(t_0) + \eta(\phi) \frac{1-C^k}{1-C} \right) e^{-\lambda(t-t_k)} \\
    &\enskip+ \eta(\phi) \\ 
    &\le K e^{-\tilde\lambda (t - t_0)} g(t_0) + \frac{K}{1-C} \eta(\phi) + \eta(\phi) \\
    &= K e^{-\tilde\lambda (t - t_0)} g(t_0) + \frac{1-C+K}{1-C} \eta(\phi)
  \end{align*}
  which coincides with~(\ref{eq:exp-2}).\qed
\end{pf}

\subsection{Sufficient condition for IPSS from ISS}
\label{sec:IPSS-from-ISS}

The following is an example of a system that is ISS but not IPSS.
\begin{prop}\label{prop:ISS_not_IPSS}
  Consider the system
  \begin{equation}\label{eq:prop_f}
    \dot{x} = -x+(1+t)g(u-|x|) \eqqcolon f_{tv}(t,x,u),
  \end{equation}
  with $f_{tv}(t,x,u)\colon \R_{\geq 0}\times \R^n \times \R^m \to \R^n$ and $g\colon \R\to\R$ such that $g(s)=0$ for all $s\leq 0$ and $g(s)=s$ for all $s>0$.
  This system satisfies Assumption~\ref{ass:standing}, is ISS and not IPSS.
\end{prop}
\begin{pf}
    The function $f_{tv}$ is locally Lipschitz and $f_{tv}(t,0,0)=0$ for all $t$, hence Assumption~\ref{ass:standing} is clearly satisfied. In \cite[Proposition 2.9]{haiman_auto19} it is shown that system \eqref{eq:prop_f} is ISS but not iISS. Hence, system \eqref{eq:prop_f} cannot be IPSS.
\end{pf}
The feature that prevents system~(\ref{eq:prop_f}) from being IPSS (and hence iISS) is that $f_{tv}$ is unbounded as a function of $t$ for some values of $(x,u)$. 
%
%
The following stronger assumptions will hence be required for ensuring IPSS of an ISS system.
\begin{assmpt}
  \label{ass:suff_ass}
  Let $f \colon \Rp \times \R^n \times \R^m \to \R^n$ satisfy \ref{item:f00} 
 of Assumption~\ref{ass:standing} and the following:
  \begin{enumerate}[label=A\arabic*)]
  \item for every $R\geq 0$ $\exists L=L_f(R)$ such that\label{item:fLip}
    \[
    |f(t,\xi_1,\mu_1)-f(t,\xi_2,\mu_2)|\leq L(|\xi_1-\xi_2|+|\mu_1-\mu_2|) \quad 
    \]
    for all $t\geq 0$ , $\xi_1,\xi_2 \in\B_R^n$, $\mu_1,\mu_2 \in \B_R^m$;
  \item there exists a zero-measure set $\cT\subset \R_{\geq 0}$ such that $f|_{\R_{\geq 0}\backslash \cT \times \R^n \times \R^m}$ is continuous.\label{item:fcont}
  \end{enumerate}
\end{assmpt}
Note that \ref{item:fcont} implies \ref{item:fmeast} and \ref{item:fLip} implies \ref{item:fcontxu}, and together with \ref{item:f00} implies that for every $R\geq 0$, there exists $M=2RL$ such that
  \begin{align}\label{eq:fboundM}
  |f(t,\xi,\mu)|\leq M \quad \forall t\geq 0, \xi\in \B_R^n, \mu \in \B_R^m.
  \end{align}
Therefore, also \ref{item:fbb} is satisfied and Assumption~\ref{ass:suff_ass} implies Assumption~\ref{ass:standing}.
%
The following is another main result.
\begin{thm}\label{thm:ISS_to_IPSS}
  Let Assumption~\ref{ass:suff_ass} hold and let system~\eqref{eq:f} be ISS. Then, system~\eqref{eq:f} has a dissipation-form ISS-Lyapunov function and is IPSS.
\end{thm}
The proof of Theorem~\ref{thm:ISS_to_IPSS} requires the consideration of a system of the form 
\begin{align}
    \label{eq:f_perturbed}
    \dot{x} &= g(t,x,d)
\end{align}
with disturbance $d$ taking values in the set 
\begin{align}
    \label{eq:cD}
    \cD &= \{d \in \cU \colon d(t) \in \B_1^m\colon \forall t\ge 0 \},
\end{align}
so that the third argument of $g$ is always bounded. 
\begin{assmpt}
    \label{ass:gpert}
    Let $g\colon \R_{\geq 0} \times \R^n \times \B_1^m \to \R^n$ satisfy
  \begin{enumerate}[label=C\arabic*)]
  \item $g(t,\cdot,\cdot)$ is continuous for all $t\geq 0$;\label{item:fbarcont}
  \item For every $R\geq 0$, there exist $M=M(R)$ and $L=L(R)$ such that\label{item:fbarLip}
    \begin{enumerate}[label=\roman*)]
    \item $|g(t,\xi,\nu)|\leq M(R)$\\
      \mbox{} \quad for all $t\geq 0$, $\xi\in\B_R^n$, $\nu\in \B_1^m$;
    \item $|g(t,\xi,\nu)-g(t,\zeta,\nu)|\leq L|\xi-\zeta|$\\
      \mbox{} \quad for all $t\geq 0$, $\xi,\zeta\in\B_R^n$, $\nu\in\B_1^m$;
    \end{enumerate}
  \item There exists a zero-measure set $\cT\subset \R_{\geq 0}$ such that $g|_{\R_{\geq 0}\backslash \cT \times \R^n \times \B_1^m}$ is continuous.\label{item:fbarcontz}
  \end{enumerate}    
\end{assmpt}

Note that \ref{item:fbarcontz} implies that $g(\cdot,\xi,\nu)$ is measurable for every fixed $(\xi,\nu)$ and that under the Lipschitz continuity on the state imposed by \ref{item:fLip} or \ref{item:fbarLip}, the solutions of \eqref{eq:f} or \eqref{eq:f_perturbed} are unique.

The proof of Theorem~\ref{thm:ISS_to_IPSS} requires the converse Lyapunov theorem to be given as Theorem~\ref{thm:converse}, for which the bounds on solutions given by the following lemma will be needed. The proof of these two results are given in Sections~\ref{sec:pf-lemma-Lip} and~\ref{sec:pf-thm-conv}.
\begin{lem}\label{lem:Lip_solution}
  Let Assumption~\ref{ass:gpert} hold and let system~(\ref{eq:f_perturbed}) be $\cD$-URLS, with $\cD$ as in~\eqref{eq:cD}. Then, for every $R,T\in\Rp$ there exists $\bar L=\bar L(R,T)\in\Rp$ such that for any $0\leq t_0\leq t\leq t_0+T$, $h\geq 0$, $\xi_1,\xi_2\in\B_R^n$ and $d\in\cD$, the unique solution $\bar x$ of~(\ref{eq:f_perturbed}) satisfies
  \begin{subequations}\label{Lip_solution}
    \begin{align}
      |\bar x(t,t_0,\xi_1,d)-\bar x(t,t_0,\xi_2,d)|\leq \bar L|\xi_1-\xi_2|,\label{Lip_solution_1}\\
      |\bar x(t+h,t_0+h,\xi_1,d)-\bar x(t,t_0,\xi_1,d)|\leq \bar Lh.\label{Lip_solution_2}
    \end{align}
  \end{subequations}	
\end{lem}

\begin{thm}\label{thm:converse}
  Let Assumption~\ref{ass:gpert} hold and suppose that the system \eqref{eq:f_perturbed} is $\cD$-URGAS, with $\cD$ as in~\eqref{eq:cD}. Then, there exists a function $V\colon \R_{\geq 0} \times \R^n \to \R$ such that the following hold:
      \begin{enumerate}[label=\alph*)]
      \item There exist $\alpha_1,\, \alpha_2 \in \cK_\infty$ such that \eqref{eq:sandwich} holds.\label{item:1}
      \item 
        For all $R\geq 0$ there exists $L=L_V(R)\geq 0$ such that $$|V(t_1,\xi_1)-V(t_2,\xi_2)|\leq L(|t_1-t_2|+|\xi_1-\xi_2|)$$ for all $t_1,t_2 \in\R_{\ge 0}$, $\xi_1,\xi_2 \in \B_R^n$.\label{item:2}
      \item For all $t \in \R_{\geq 0}\backslash \cT$, all $\xi \in \R^n$ and all $\nu\in\B_1^m$
	\begin{align*}
	  D^+_{g} V(t,\xi,\nu)
	  \leq -\frac{V(t,\xi)}{2}.
	\end{align*}\label{item:3}
      \end{enumerate}
\end{thm}
\hernan{The converse Lyapunov result in \cite[Theorem~3.5]{karjia_book11} requires different, in general weaker, assumptions and establishes the existence of a locally Lipschitz function $V:\R_{\ge 0} \times \R^n \to \R^n$. By contrast, the function $V$ of Theorem~\ref{thm:converse} is ensured to have the stronger property of being Lipschitz on $\R_{\ge 0} \times \B_R^n$ for every $R\ge 0$. The latter uniformity of the Lipschitz constant for all $t\ge 0$ is an essential requirement in the proof of Theorem~\ref{thm:ISS_to_IPSS}, which is next developed.}

\textbf{PROOF of Theorem~\ref{thm:ISS_to_IPSS}.}
Since~\eqref{eq:f} is ISS (uniformly with respect to initial time), then following the proof of \cite[Lemma~2.12]{sonwan_scl95}, there exists a function\footnote{A $\cC^\infty$ function $\varphi$ actually exists, but $\cC^1$ suffices here.} $\varphi\in \cC^1(\R_{\geq 0})\cap \cK_\infty$ such that system~\eqref{eq:f_perturbed} with
	\begin{align}
	   g(t,\xi,\nu) = f(t,\xi,\nu\varphi(|\xi|)), 
	\end{align}
	is $\cD$-URGAS. Since $f$ satisfies Assumption~\ref{ass:suff_ass} and $\varphi \in \cC^1$, then $g$ satisfies Assumption~\ref{ass:gpert}. Let $V : \Rp \times \R^n \to \Rp$ be the function given by Theorem~\ref{thm:converse}. Then,
	\[ D^+_{g} V(t,\xi,\nu) 
        \leq -\frac{V(t,\xi)}{2}
	\]
	for all $t \in \R_{\geq 0}\backslash \cT$, $\xi\in\R^n$ and $\nu$ such that $|\nu|\le 1$. This means that 
        \begin{align}\label{eq:DpV-case1}
          D^+_{f} V(t,\xi,\mu) \le -\frac{V(t,\xi)}{2} \quad \text{if }|\xi|\geq \varphi^{-1}(|\mu|),
        \end{align}
         or equivalently if $|\mu| \leq \varphi(|\xi|)$.
	
	Let $L_V(R)$ and $L_f(R)$ be the Lipschitz constants of $V$ and $f$, 
    as given by Theorem~\ref{thm:converse} and Assumption~\ref{ass:suff_ass}. Without loss of generality, assume that $L_V$, $L_f$ are continuous and increasing functions. If $t\in \R_{\geq 0}\backslash\cT$ and $|\xi|\leq \varphi^{-1}(|\mu|)$ we have that
	\begin{multline}
  \frac{V(t+h,\xi+hf(t,\xi,\mu))-V(t,\xi)}{h} = \\
		= \frac{V(t+h,\xi+hf(t,\xi,\mu))-V(t+h,\xi+hf(t,\xi,0))}{h} \\
		+\frac{V(t+h,\xi+hf(t,\xi,0))-V(t,\xi)}{h} .\label{eq:DpV1}
	\end{multline}
	By Theorem~\ref{thm:converse}, the last term above satisfies 
    $$ \limsup_{h \to 0^+} \frac{V(t+h,\xi+hf(t,\xi,0))-V(t,\xi)}{h} \leq -\frac{V(t,\xi)}{2}.$$
    As for the first term of~\eqref{eq:DpV1}, if $|h|\leq 1$, then 
    \begin{align*}
        |\xi&+hf(t,\xi,\mu)|\leq |\xi|+|f(t,\xi,\mu)|\\
        &\leq |\xi|+L_f(\max\{|\xi|,|\mu|\})[|\xi|+|\mu|]\\
        &\leq \varphi^{-1}(|\mu|)+L_f(\max\{\varphi^{-1}(|\mu|),|\mu|\})[\varphi^{-1}(|\mu|)+|\mu|]\\
        &\coloneqq \chi_1(|\mu|),\\[1mm]
        \lefteqn{\text{and}\quad h|f(t,\xi,\mu)-f(t,\xi,0)|\leq}\phantom{|\xi} \\
        &\qquad L_f(\max\{\varphi^{-1}(|\mu|),|\mu|\})|\mu|
        :=\chi_2(|\mu|).
    \end{align*}
    Through the Lipschitz constants of $V$ and $f$, it follows that
	\begin{multline*}
		\left\lvert \frac{V(t+h,\xi+hf(t,\xi,\mu))-V(t+h,\xi+hf(t,\xi,0))}{h} \right\rvert \\
		\leq L_V(\chi_1(|\mu|))\chi_2(|\mu|) \coloneqq \tilde{\rho}(|\mu|),
	\end{multline*}
	with $\tilde{\rho}\in\cK_\infty$. 
	Taking the two terms into account,
    \begin{align}\label{eq:DpV-case2}
        D^+_f {\scriptstyle V(t,\xi,\mu)} \leq -\frac{\scriptstyle V(t,\xi)}{2} + \tilde{\rho}(|\mu|) \quad \text{if $|\xi|\leq \varphi^{-1}(|\mu|)$.}
    \end{align}
    Combining~\eqref{eq:DpV-case1} and~\eqref{eq:DpV-case2}, 
    then for all $t \in \R_{\geq 0} \backslash \cT$, $\xi \in \R^n$ and $\mu\in\R^m$,
	\begin{align*}
        D^+_f V(t,\xi,\mu) 
		\leq -\frac{V(t,\xi)}{2} + \tilde{\rho}(|\mu|).
	\end{align*}
    This establishes that $V$ is a dissipation-form ISS-Lyapunov function for system~\eqref{eq:f}. Since Assumption~\ref{ass:suff_ass} implies Assumption~\ref{ass:standing}, then application of Theorem~\ref{thm:Lyap-ipss} establishes that system~\eqref{eq:f} is IPSS.
\qed

\hernan{In view of Lemma~\ref{lem:ipss-stronger} and Theorems~\ref{thm:Lyap-ipss} and~\ref{thm:ISS_to_IPSS}, the question of whether IPSS could actually be \emph{equivalent} to the combination of ISS and iISS remains unanswered only under assumptions weaker than Assumption~\ref{ass:suff_ass}.}

\subsection{Proof of Lemma~\ref{lem:Lip_solution}}\label{sec:pf-lemma-Lip}
    Let $R,T\in\Rp$ be arbitrary and define $R_1=\varepsilon(R)$ and $R_2=\varepsilon(R_1)$, with $\varepsilon(\cdot)$ the function characterizing the $\cD$-URLS property.
  Note that $R_2\geq R_1\geq R$.
  Set $L_2:=L(R_2), M_1:=M(R_1)$, with $L,M$ from \ref{item:fbarLip} in Assumption~\ref{ass:gpert}.
  For all $0\leq t_0\leq t$, $\xi\in\B_{R_1}^n$ and $d\in\cD$, we have
  \begin{equation*}
    |\bar x(t,t_0,\xi,d)|\leq 
    \varepsilon(R_1) = R_2.
  \end{equation*}
  Let $\xi_i\in\B_{R_1}^n$ and denote $\bar x_i(\cdot):=\bar x(\cdot,t_0,\xi_i,d)$ for $i=1,2$. Note that for any $t\geq t_0$, $\bar x_i(t)=\bar x(t,t_0,\xi_i,d)=\xi_i+\int_{t_0}^tg(s,\bar x_i(s),d(s))ds$. Therefore by triangle inequality and the Lipschitz condition on $g$, we have
  \begin{align*}
    |\bar x_1(t) &-\bar x_2(t)|\leq |\xi_1-\xi_2|\\
    &+\int_{t_0}^t\left|g(s,\bar x_1(s),d(s))-g(s,\bar x_2(s),d(s))\right|ds\\
    &\leq|\xi_1-\xi_2|+\int_{t_0}^t L_2|\bar x_1(s)-\bar x_2(s)|ds.
  \end{align*}
  It follows from Gronwall's inequality that if ${t\leq t_0+T}$, 
  \begin{equation}\label{step_A}
    |\bar x_1(t)-\bar x_2(t)|\leq |\xi_1-\xi_2|e^{L_2(t-t_0)}\leq |\xi_1-\xi_2|e^{L_2T}.
  \end{equation}
  Next, for all $0\leq t_0\leq t\leq t_0+T, h\geq 0, \xi\in\B_R^n$, $d\in\cD$, it holds that
  \begin{align}\label{step_B}
    &|\bar x(t+h,t_0+h,\xi,d)-\bar x(t,t_0,\xi,d)|\nonumber \\
    &\leq|\bar x(t+h,t_0,\xi,d)-\bar x(t,t_0,\xi,d)|\nonumber\\
    &\quad+|\bar x(t+h,t_0+h,\xi,d)-\bar x(t+h,t_0,\xi,d)|.
  \end{align}	
  Denote $\bar x(\cdot):=\bar x(\cdot,t_0,\xi,d)$.
  On the one hand, 
  \begin{multline}\label{step_C}
    |\bar x(t+h,t_0,\xi,d)-\bar x(t,t_0,\xi,d)|=|\bar x(t+h)-\bar x(t)|\\
    =\left|\int_t^{t+h}g(s,\bar x(s),d(s))ds\right| \leq \int_t^{t+h}M_1ds=M_1h.
  \end{multline}
  On the other hand, note that $\bar x(t+h,t_0,\xi,d)=\bar x(t+h,t_0+h,\bar x(t_0+h),d)$. Since both $\xi,\bar x(t_0+h)\in\B_{R_1}^n$, and $t_0+h\leq t+h\leq t_0+h+T$, it follows from \eqref{step_A} that
  \begin{multline}\label{step_D}
    |\bar x(t+h,t_0+h,\xi,d)-\bar x(t+h,t_0,\xi,d)|\\
    =|\bar x(t+h,t_0+h,\xi,d)-\bar x(t+h,t_0+h,\bar x(t_0+h),u)|\\
    \leq|\xi-\bar x(t_0+h)|e^{L_2T}\leq M_1e^{L_2T}h,
  \end{multline}
  where the second inequality follows from~(\ref{step_C}) with $t$ replaced by $t_0$.
  Substituting \eqref{step_C} and \eqref{step_D} into \eqref{step_B}, we conclude
  \begin{equation*}
    |\bar x(t+h,t_0+h,\xi,d)-\bar x(t,t_0,\xi,d)|\leq M_1(1+e^{L_2T})h,
  \end{equation*}
  which, together with \eqref{step_A}, implies \eqref{Lip_solution} with $\bar L=\max\{e^{L_2T},M_1(1+e^{L_2T})\}$ for all $0\leq t_0\leq t\leq t_0+T, \xi_1,\xi_2\in\B_R^n$ and $d\in\cD$.\qed

\subsection{Proof of Theorem~\ref{thm:converse}}\label{sec:pf-thm-conv}
The proof of this theorem follows the lines of the proof of \cite[Theorem~B.31]{ISS:Mironchenko:2023} for time-invariant systems.  
  Under the given assumptions, for every $(t_0,\xi_0,d) \in \R_{\ge 0} \times \R^n \times \cD$ there exists a unique maximally defined and forward complete solution of (\ref{eq:f_perturbed}), denoted by $\bar x(t,t_0,\xi_0,d)$, that satisfies $\bar x(t_0,t_0,\xi_0,d) = \xi_0$. Let 
  $\beta\in\KL$ be the corresponding function for which (\ref{eq:RGAS}) holds with $x$ replaced by $\bar{x}$. Due to Sontag's Lemma~\cite[Proposition 7]{ISS:Sontag:1998}, there exist $\theta_1,\theta_2\in\cK_\infty$ so that
  \begin{equation}
    \theta_2^{-1}(\beta(s,t))\leq\theta_1(s)e^{-t},\quad\forall s,t\in\Rp.
  \end{equation}
  Define $\rho: \Rp \to \Rp$ as
  \begin{equation}
    \rho(s):= \inf_{r\ge 0} \left\{\theta_2^{-1}(r) + |r-s|\right\}. 
  \end{equation}
  It follows from\footnote{A similar construction is given in \cite[Lemmas 7, 8]{Liu2021b}.} \cite[Lemma~A.18]{ISS:Mironchenko:2023} that $\rho\in\cK_\infty$,   $\rho(s)\leq\theta_2^{-1}(s)$ for all $s\in\Rp$ and $\rho$ is globally Lipschitz with unit Lipschitz constant. For any $k\in\N$, define $G_k(r):=\max\{r-\frac{1}{k},0\}$ and consider a function
  \begin{equation*}
    W_k(t_0,\xi):=\sup_{d\in\cD}\sup_{s\geq t_0}e^{\frac{1}{2}(s-t_0)}G_k\big(\rho(|\bar x(s,t_0,\xi,d)|)\big),
  \end{equation*}
  for all $t_0\ge 0, \xi\in\R^n$. 
  By the arguments in \cite[Theorem~B.31]{ISS:Mironchenko:2023}, it follows that
 \begin{align*}
    0\leq W_k(t_0,\xi) \leq \theta_1(|\xi|).
  \end{align*}
  Thus $W_k$ is bounded from above. 
	
  Next, we show that $W_k$ is Lipschitz on $\Rp\times\B_R^m$ for all $R\in\Rp$. Pick any $R\geq 0$. Define $T_{R,k}:=\ln(1+k\theta_1(R))$. Since the system is $\cD$-URGAS, for any $t_0\geq 0, t\geq t_0+T_{R,k}$, $\xi\in\B_R^n$ and $d\in\cD$, 
  \begin{align*}
    \rho(|\bar x &(t,t_0,\xi,d)|)
    \leq \theta_2^{-1}\left(\beta(|\xi|,t-t_0)\right)
    \leq e^{-(t-t_0)}\theta_1(|\xi|) \\
    &\leq e^{-T_{R,k}}\theta_1(R) 
    = \frac{\theta_1(R)}{1+k\theta_1(R)}<\frac{1}{k}
  \end{align*}
  and it follows from the definition of $G_k$ that
  \begin{equation*}
    G_k\big(\rho(\bar x(t,t_0,\xi,d))\big)=0 .
  \end{equation*}
  From this and continuity, then the inner supremum in the definition of $W_k$ is a maximum, attained within the interval $[t_0, t_0+T_{R,k}]$. Therefore, for any $t_0\geq 0, \xi\in\B_R^n$,
  \begin{align*}
    W_k(t_0,\xi) 
    &=\sup_{d\in\cD}\max_{s\in [0,T_{R,k}]}e^{\frac{s}{2}}G_k\big(\rho(|\bar x(s+t_0,t_0,\xi,d)|)\big).
  \end{align*}	
  Pick any $\xi,\zeta\in\B_R^n$ and $t,p\geq 0$. We have
  \begin{align*}
    &|W_k(t,\xi)-W_k(p,\zeta)|\\
    &\leq \sup_{d\in\cD}\max_{s\in [0,T_{R,k}]} \Big[ e^{\frac{1}{2}s}\Big|G_k\big(\rho(|\bar  x(s+t,t,\xi,d)|)\big)\\
      &\hspace{32mm} -G_k\big(\rho(|\bar x(s+p,p,\zeta,d)|)\big)\Big| \Big]\\
    &\leq \sup_{d\in\cD}\max_{s\in [0,T_{R,k}]}e^{\frac{1}{2}s}\big||\bar x(s+t,t,\xi,d)|-|\bar x(s+p,p,\zeta,d)|\big|\\
    &\leq \sup_{d\in\cD}e^{\frac{T_{R,k}}{2}}\max_{s\in [0,T_{R,k}]}\big|\bar x(s+t,t,\xi,d)-\bar x(s+p,p,\zeta,d)\big|
  \end{align*}
  where we have used \cite[Lemma B.29]{ISS:Mironchenko:2023} for the first inequality and the fact that both $G_k,\rho$ are Lipschitz with unit Lipschitz constant for the second inequality. Add and subtract $\bar x(s+p,p,\xi,d)$ within the norm, note that $\cD$-URGAS implies $\cD$-URLS, and employ Lemma~\ref{lem:Lip_solution},
  \begin{align}
    \label{eq:WkLip}
    |W_k(t,\xi)-W_k(p,\zeta)| &\le M_{R,k} (|t-p|+|\xi-\zeta|),\\
    \label{eq:MRkdef}
    M_{R,k} &:= e^{\frac{T_{R,k}}{2}}\bar L(R,T_{R,k}).
  \end{align}
  This ensures that $W_k$ is Lipschitz on $\Rp\times\B_R^m$ for all $R\ge 0$, with Lipschitz constant $M_{R,k}$. Note that $M_{R,k}$ is a strictly increasing function with respect to both arguments.
  Next, define
  \begin{equation}
    \label{eq:defV}
    V(t,\xi):=\sum_{k=1}^{\infty}\frac{2^{-k}}{1+M_{k,k}}W_k(t,\xi).
  \end{equation}
  We claim that this $V$ is a desired Lyapunov function for the system \eqref{eq:f_perturbed}, which satisfies all the conditions in Theorem~\ref{thm:converse}.
	
  Since for all $t\ge 0, \xi\in\R^n, k\in\N$, it holds that $W_k(t,\xi)\leq \theta_1(|\xi|)$, and
  \begin{equation*}
    W_k(t,\xi)\geq \sup_{d\in\cD}G_k\big(\rho(|\bar{x}(t,t,\xi,d)|)\big)=G_k(\rho(|\xi|)),
  \end{equation*}
  we conclude that
  \begin{equation}
    \alpha_1(|\xi|)\leq V(t,\xi)\leq \alpha_2(|\xi|),
  \end{equation}
  where
  \begin{equation*}
    \alpha_1(r):=\sum_{k=1}^{\infty}\frac{2^{-k}}{1+M_{k,k}}G_k(\rho(r)),
  \end{equation*}
  and $\alpha_2:=\theta_1\in\cK_\infty$. The fact that $\alpha_1\in\cK_\infty$ is established in the proof of \cite[Theorem B.31]{ISS:Mironchenko:2023}. This verifies item \ref{item:1}.
	
	Next, for any $R\ge 0$, $\xi,\xi'\in\B_R^n$ and $t,t'\geq 0$, then
	\begin{align*}
		|V(t,\xi)&-V(t',\xi')|\\
		&=\left|\sum_{k=1}^{\infty}\frac{2^{-k}}{1+M_{k,k}}\big(W_k(t,\xi)-W_k(t',\xi')\big)\right|\\
		&\leq \sum_{k=1}^{\infty}\frac{2^{-k}M_{R,k}}{1+M_{k,k}}\left(|t-t'|+|\xi-\xi'|\right)\\
		&\leq \left(1+\sum_{k=1}^{\lfloor R\rfloor+1}\frac{2^{-k}M_{R,k}}{1+M_{k,k}}\right)\left(|t-t'|+|\xi-\xi'|\right),
	\end{align*}
    where the last inequality holds due to the fact that $M_{R,k}\le M_{k,k}$ when $k\ge \lfloor R\rfloor+1$. This verifies item \ref{item:2} with $L(R):=1+\sum_{k=1}^{\lfloor R\rfloor+1}\frac{2^{-k}M_{R,k}}{1+M_{k,k}}$.
 
Let $t_0\in \Rp\setminus \cT$, $\xi\in \R^n$ and $\nu \in \B_1^m$. Define $c\in\cD$ via $c(t)\equiv \nu$ and consider $\bar{x}=\bar x(\cdot,t_0,\xi_0,c)$. From \ref{item:fbarcontz} of Assumption \ref{ass:gpert}, it follows that $\dot{\bar x}(t_0)=g(t_0,\xi_0,\nu)$. Define $v(t)=V(t,\bar x(t))$ and $w_k(t) := W_k(t,\bar x(t))$ for $t\ge t_0$ and $k\in \N$. 
  Observe that for 
  $t\geq t_0$,	
  \begin{align*}
    &w_k(t) 
    =\sup_{d\in\cD}\sup_{s\geq t}e^{\frac{1}{2}(s-t)}G_k\big(\rho(|\bar x(s,t,\bar x(t),d)|)\big)\\
    &=e^{-\frac{1}{2}(t-t_0)}\sup_{d\in\cD}\sup_{s\geq t}e^{\frac{1}{2}(s-t_0)}G_k\big(\rho(|\bar x(s,t_0,\xi,c\conc{t}d)|)\big),
  \end{align*}
  where $c\conc{t} d$ is defined in Section~\ref{sec:notation}.
  Since $c\conc{t} d\in\cD$, we further conclude that 
  \begin{align*}
    w_k(t) 
    &\le e^{-\frac{1}{2}(t-t_0)}W_k(t_0,\xi) = e^{-\frac{1}{2}(t-t_0)}w_k(t_0).
  \end{align*}
  Then, if $h>0$
  \begin{align*}
  &\frac{v(t_0+h)-v(t_0)}{h}=\sum_{k=1}^{\infty}\frac{2^{-k}}{1+M_{k,k}}\left(\frac{w_k(t+h)-w_k(t_0)}{h}\right )\\
  &\le \frac{e^{-\frac{1}{2}h}-1}{h} \left (\sum_{k=1}^{\infty}\frac{2^{-k}}{1+M_{k,k}} w_k(t_0)\right ) 
  =\frac{e^{-\frac{1}{2}h}-1}{h} v(t_0).
  \end{align*}
  Therefore
  \begin{align*}
  \limsup_{h\to 0^+} \frac{v(t_0+h)-v(t_0)}{h}\le -\frac{v(t_0)}{2}.
  \end{align*}
  Thus, 
  from the Lipschitz property of $V$ and the fact that $\dot{\bar x}(t_0)=g(t_0,\xi_0,\nu)$ , it follows that for all $t_0\in \Rp\setminus \cT$, $\xi \in \R^n$ and $\nu \in \B_1^m$
  \begin{align*}
    D^+_{g} V(t_0,\xi,\nu)&=\limsup_{h\to 0^+} \frac{v(t_0+h)-v(t_0)}{h} 
    \leq -\frac{V(t_0,\xi)}{2},
  \end{align*}	
and item \ref{item:3} follows. This completes the proof of Theorem~\ref{thm:converse}.\qed




\section{Conclusion}

The concept of input-power-to-state stability (IPSS) was introduced and analyzed for time-varying systems under weak assumptions. This concept provides useful bounds on the state when the input has bounded average power but no a priori bounds on amplitude or energy. IPSS was shown to be not weaker than the combination of ISS and iISS, and different necessary and sufficient conditions for IPSS to hold were given. Specifically, it was shown that (i) IPSS is implied by the existence of a dissipation-form ISS-Lyapunov function, (ii) IPSS is implied by an iISS bound on the state norm with an exponential class-$\KL$ function, (iii) ISS under stronger assumptions related to uniform over time Lipschitz continuity and boundedness implies the existence of a dissipation-form ISS-Lyapunov function and hence IPSS. An interesting question for future research is whether IPSS is in fact equivalent to the existence of a dissipation-form ISS-Lyapunov function under weaker assumptions.

\bibliographystyle{plain}
\bibliography{references.bib}

\end{document}